\documentclass[12pt]{spieman}  
\usepackage{amsmath,amsfonts,amssymb}
\usepackage{graphicx}
\usepackage{setspace}
\usepackage{tocloft}
\usepackage{lineno} 

\title{Understanding Brain Aging Across Populations: A Comprehensive Framework for Structural Analysis}

\author[a,*]{Alphin J Thottupattu}
\author[a]{Jayanthi Sivaswamy}
\author[b]{Bharath Holla}
\author[b]{Jithender Saini}
\affil[a]{Centre for Visual Information Technology (CVIT), International Institute of Information Technology, Street,
Hyderabad, 50032, Telangana, India}
\affil[b]{Department of Psychiatry, National Institute of Mental Health and Neurosciences, Bangalore, Karnataka, India}

\cftpagenumbersoff{figure}
\cftpagenumbersoff{table} 
\begin{document} 
    \maketitle

\begin{abstract}
\section*{Significance}
Understanding distinct neurological aging patterns across various populations is vital in the context of a globally aging populace. This study is motivated by the necessity to explore the structural variations in brain aging across different populations.

\section*{Aim} Our aim is to develop an elaborate framework for comparing structural variations across aging in diverse populations and to demonstrate the applicability of the proposed framework with a sample MRI data from Indian, Chinese, Japanese, and Caucasian groups.

\section*{Approach}
We introduce an elaborate framework for comparing structural variations across aging in diverse populations using MRI data. The analysis involves a two-pronged approach. Firstly, a group analysis is conducted involving tissue segmentation through FSL-FAST, examining gray matter (GM), white matter (WM), and cerebrospinal fluid (CSF). Subsequently, a continuous model-based analysis is employed, defining aging as a diffeomorphic transformation. This approach facilitates a detailed intra- and inter-population analysis, examining both global anatomy and age-dependent distances of each population in comparison to the Indian population.

\section*{Results}
This framework provides qualitative and quantitative insights into aging-related structural changes. Findings show distinct aging trajectories across populations, including differences in the onset of gray matter reductions and ventricular expansion. Hemispherical asymmetry in left brain CSF-filled region expansion is significant across all populations. Furthermore, detailed insights into the spatial distribution of brain deformations over time, with a particular focus on anatomical changes relative to a reference time point, are provided. This facilitates direct comparison of structural changes across populations.

\section*{Conclusion}
The proposed comprehensive structural comparison framework, applied to a sample dataset encompassing four distinct populations, represents a pioneering effort to compare structural differences related to brain aging on a global scale. Subsequent studies utilizing larger datasets and applying the proposed analysis across groups based on various criteria will further advance our understanding of aging-related changes.
\end{abstract}

\keywords{Structural changes, Aging, Cross-population Analysis}

{\noindent \footnotesize\textbf{*}Alphin J Thottupattu,  \linkable{alphinj.thottupattu@research.iiit.ac.in}} 

\begin{spacing}{2}   

\section{Introduction}
Brain aging encompasses a wide array of changes in brain morphology and cognitive abilities. Understanding the standard patterns of aging is essential for identifying normal and abnormal trends. Although these changes vary among individuals, studying them within a cohort can reveal general aging trends influenced by factors such as gender, educational background, and multilingualism \cite{age_sex, age_edu, age_language}. The focus of this study is specifically on comparing the structural changes in healthy brain aging across different populations while excluding cognitive changes.\\
Promoting healthy aging involves recognizing standardized expectations and detecting deviations from them. For instance, research has shown that aerobic exercise \cite{strategy} positively impacts brain aging, while engaging in activities like multilingualism and music \cite{multilug} may also contribute to healthy aging. To create effective strategies for healthy aging, it is vital to grasp the trends and deviations in aging, particularly among diverse populations. Understanding why certain populations exhibit better aging outcomes will enable the implementation of methods that promote better overall aging outcomes for everyone.\\
Progress in understanding brain aging has been significant, with global standards established through the analysis of diverse brain scans \cite{chart}. Various methods, such as age-specific templates \cite{AGE_SEPC_ATLAS1_brains,AGE_SEPC_ATLAS2_Richard} and spatiotemporal models \cite{ours,huizinga2018spatio}, have been utilized to explore population-specific structural changes during aging. While template-based techniques offer insights into long-term anatomical changes, spatiotemporal models capture the dynamic nature of aging, including changes in gross structure, iron deposition, and demyelination \cite{common_temp_space1}. However, these studies have been limited to single populations. Global data-driven studies, exemplified by \cite{chart}, focus on universal patterns using extensive datasets. Considering factors like gender, ethnicity, and education is crucial, and studying populations with shared backgrounds can offer targeted insights compared to global studies \cite{chart}. Despite this, variations in aging trends may still exist within these populations.\\
An essential aspect of studying aging involves comprehending the variations in aging, alongside the previously discussed modeling of common trends. This is imperative because it allows for a more comprehensive understanding of the aging process. Comparing aging across populations based on different criteria facilitates this deeper understanding. One practically possible criterion is geographic locations, which are distinct in ethnicity and other factors controlling aging.
\\
Numerous studies have sought to explore and distinguish aging patterns among diverse populations. Typically, researchers analyze aging patterns by visually comparing time series data or using derived measures, such as changes in cerebral volume and brain dimensions \cite{Brain_dev, INTERESTING1, brain_dev3}. A widely employed technique for comparing population trends in brain structure is Voxel-based morphometry (VBM) \cite{comparison1, comparison2, comparison3}. VBM entails mapping brain images from different individuals to a common template space and subsequently comparing the derived cortical features across these mapped brains. However, VBM analysis does have limitations when it comes to direct population comparisons for age-related brain structure changes since it primarily focuses on detecting differences in brain structure between predefined groups, such as young versus old or patient versus control. Consequently, its main emphasis is on identifying group disparities, rather than providing a detailed understanding of age-related changes that may occur in a continuous and subtle manner across diverse populations. As such, while VBM can be valuable for exploring structural differences within specific groups, it may not be fully tailored for comprehensive investigations of brain aging across different populations.
\\In order to comprehensively understand and compare brain aging, relying solely on qualitative analysis or quantitative comparison studies based on derived features is insufficient. The complexities of brain aging require a framework that combines both qualitative and quantitative analysis methods. In this paper, we are proposing an analysis framework to compare the aging process across populations, and we demonstrate this with sample data from four different populations: Indian, Chinese, Korean, and Caucasian. These populations were chosen for their diverse backgrounds, expected to contribute to variations in the aging process. By examining brain aging in these specific populations, we seek valuable insights into unique aging characteristics within and across groups.\\
Brain aging is not solely governed by time, as various factors such as genetics, environment, and lifestyle play significant roles in influencing the aging process. In light of this complexity, our analysis aims to distinguish aging-dependent variations from the impacts of other aging-independent factors. We do this to achieve a more meaningful comparison of brain aging across different populations.
\\
In summary, our contributions include qualitative and quantitative analysis of brain aging within and across populations, distinguishing aging-dependent changes as individuals age, and age-independent differences in brain aging patterns among different groups. Most importantly, we demonstrate the applicability of our proposed method to compare aging across four populations, a comparison that has not been extensively explored in the existing literature.
\section{Data collection for framework demonstration}
Our study considered four distinct populations (Indian, Caucasian, Chinese, and Japanese) to gain insights into the trends in changes in brain anatomy with age. T1 MRI scans (data) of the brain were sourced from healthy adults aged 20 to 80 years, in each population as summarized in Table 1. The data for the Indian population were partly collected explicitly for this study and partly retrieved from previous studies. The data from previous studies \cite{holla2020}were selected to approximate the acquired data closely. The data for other populations were sourced from publicly available repositories. Ideally, we need to have equal number of subjects, both male and female, at each time point (decade in our case) in the study to avoid any bias. The number of subjects considered for each time point was constrained by the fact that the public repositories had fewer elderly than young/middle-aged subjects. Hence, we chose 26 subjects (split equally between males and females) for each time point. To maintain consistency, a similar distribution of age and gender was maintained across all populations in each decade as seen in Table 1. 
\begin{table}[ht!]
\centering
\begin{tabular}{|c|c|c|c|c|c|}
\hline
Population & \begin{tabular}[c]{@{}c@{}}Magnetic field\\ \\ strength \\ of scanners\end{tabular} & \begin{tabular}[c]{@{}c@{}}\#Subjects \\ distribution\\ in every decade\\ (20-80 years)\end{tabular} & \begin{tabular}[c]{@{}c@{}}Gender \\ Distribution\\ M:F\end{tabular} & Matrix       & Size (mm)   \\ \hline
Indian     & 3T Philips                                                                          & 1:1:0.9:1:1:1                                                                                        & 1:0.99                                                                & 256x256x165  & 0.7×0.7×0.5 \\ \hline
Caussian   & 3T Siemens                                           l0.                              & 1:1:1:1:1                                                                                            & 1:1                                                                  & 256x240x192  & 1x1x1       \\ \hline
Chinese    & 3T Siemens                                                                          & 1:1:1:1:1:0.88                                                                                       & 0.99:1                                                               & 256 x256x176 & 1x1x1       \\ \hline
Japanese   & 0.5 T GE                                                                            & 1:1:1:1:1:1                                                                                          & 1:1                                                                  & 256×256×124  & 1×1×1.5     \\ \hline
\end{tabular}
\label{datap}
\caption{Comparison of Population Characteristics and MRI Scanner Specifications for Different Ethnic Groups}
\end{table}
A detailed description of the data collected for each population is presented next.

\subsection{Indian population}
\begin{figure}[ht!]
    \centering
    \includegraphics[width=\linewidth]{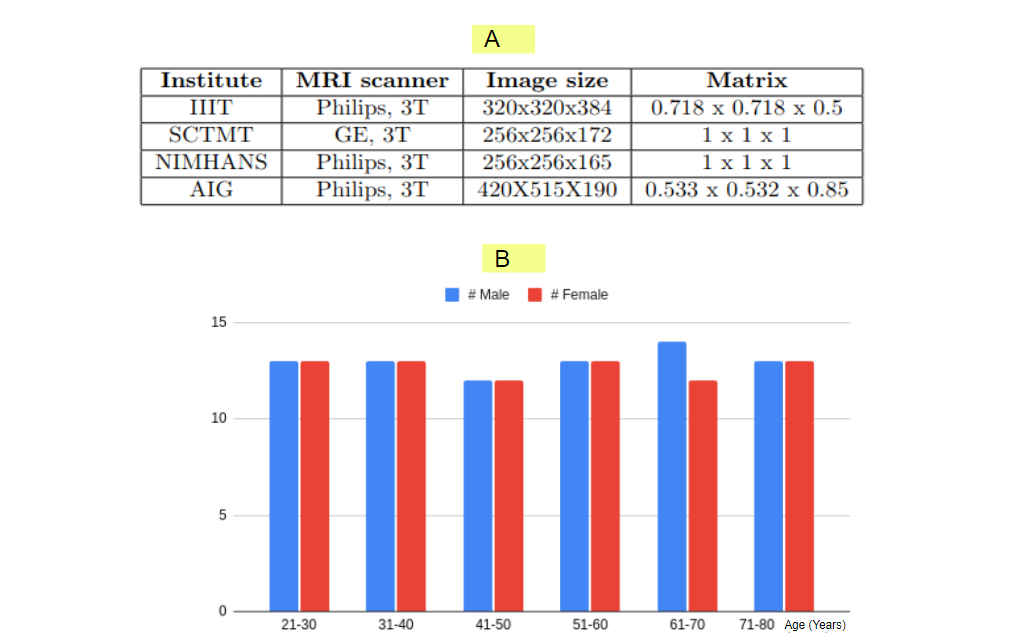}
    \caption{A) Acquisition details for the collected scans. B) Gender distribution of subjects in the Indian cohort.}
    \label{fig1}
\end{figure}

3T scans of healthy subjects were collected from multiple sites (see the table in Figure \ref{fig1}A for details) within India \cite{holla2020} and individual sites had ethical approval for collecting and sharing the data for research purposes. In the 20-80 age range, 26 scans were collected for each decade, with the exception of a single decade which had 24 scans. This dataset will be referred to as the IBASD.
The gender and age distribution of the collected data is shown in Figure \ref{fig1} B. 
Only physically and psychologically normal subjects with no history of head injury or other neurological disorders were included in the study. Pregnant women, subjects born pre-maturely, and those with any long-term disease condition were excluded from the study. All the acquired scans were of right-handed subjects. Experienced medical experts checked all the scans in corresponding institutions for any structural abnormality in the scans. 

\subsection{Caucasian population}
The Cambridge Centre for Ageing and Neuroscience (Cam-CAN) database \cite{camcan} has neuroscans of subjects recruited from the Cambridge City area in the United Kingdom. A decadal of 26, 3T MRI scans were selected for each decade from the Cam-CAN database for our study. The demographic information available in the database for each individual was used to select subjects labeled as belonging to the 'white' ethnic group. 

\subsection{Chinese population}
The Chinese population data was sourced from the Southwest University Adult Lifespan Dataset (SALD) \cite{sald} built with scans of University students, staff, and subjects from control group of clinical studies. Out of 494 scans available, 153 scans within the age range of 20-80 years were carefully selected to maintain gender distribution and an equal number of subjects in each decade. A total of 23 images were selected in the 70-80 age range, with rough gender balance, while 26 images were selected in other age ranges.

\subsection{Japanese population}
The Japanese population data was sourced from the AOBA database \cite{aoba} which includes scans of volunteers recruited by the AOBA Brain Imaging Research Center in Japan. Out of a total of 1153 scans available in the database, 156 scans were chosen for our study which resulted in 26 scans for each decade. Although the scans were sourced from a 0.5T scanner, they were of good quality and comparable to other databases, possibly due to longer scanning times.

\section{Methodology}
This study obtained ethical approval from an expert committee, ensuring that informed consent was acquired from all participants whose MRIs are utilized in this research. All methods were performed in accordance with the relevant guidelines and regulations. Additionally, data sourced from other studies received a separate ethical clearance specifically for inclusion in this study. All processing of such data was conducted at the corresponding sites in the presence of the respective data owners.\\
A time series of templates developed from the collected data representing each decade forms the basis of our aging study. Given that the scans we selected were collected from multiple sites with different scanners and protocols, the scans have  to be standardised and normalised. 
\subsection{ Data Normalization and Standardization}
We employed two analysis strategies to compare the aging of different populations: group analysis and model-based analysis. The data preparation process to support these two analysis strategies was different. The specific processes are explained here.
\subsubsection{Preprocessing for Group analysis} 
Individual images within all population-specific datasets (CAMCAN, SALD, AOBA, and IBASD) were preprocessed using the complete FSL pipeline, including the "$fsl\_anat$" command. This comprehensive pipeline involves various steps such as brain extraction, bias field correction, tissue segmentation, and nonlinear registration. This preprocessing pipeline ensures consistent and reliable quality in tissue segmentation, which is critical for group-level analysis and comparisons.
\subsubsection{Data Preparation for Aging Model Based analysis}
\begin{figure}[ht!]
    \centering
    \includegraphics[width=\linewidth]{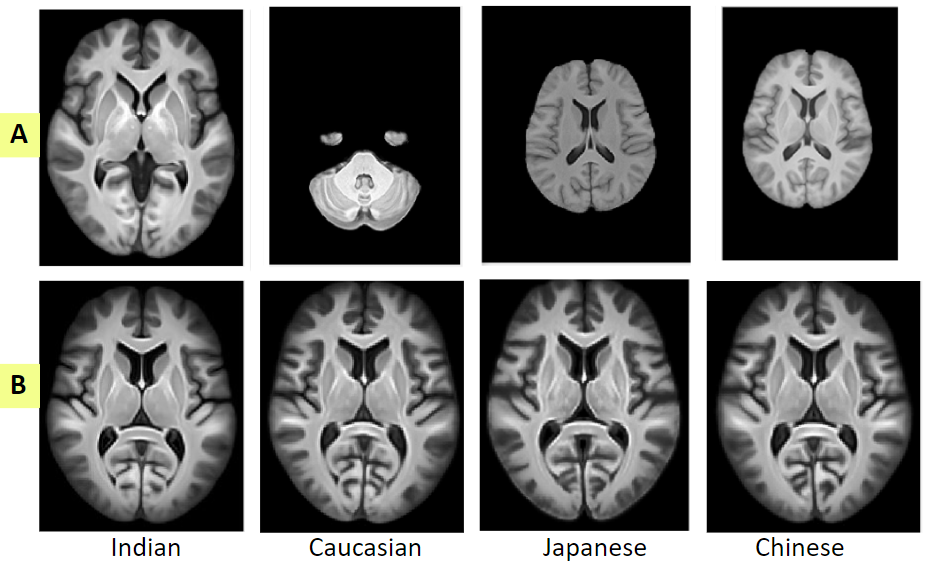}
    \caption{ Sample templates corresponds to 40-50 years age range before(A) and after (B) template pre-processing }
    \label{BB}
\end{figure}

The main goal of model-based analysis is to develop a consistent and generalizable model that can be applied across datasets, rather than performing analysis on derived measurements from individual images.\\
\textit{Creating Decade-Specific Templates} -
For each population, individual images of subjects, in the age range of 20 to 80 years, were skull-stripped and grouped into six distinct decades, for each population. The group registration tool from the ANTS package was then used to create templates for each decade. The image dimension, intensity range, and alignment were maintained relative to the individual population space, as illustrated in Figure \ref{BB}-A. However, it is important to note that the templates needed to be aligned to a common reference space before proceeding to the subsequent steps, which will be discussed in detail later on.\\
\textit{Creating Reference Template} -  A reference template was crafted to align all templates to a common space, closely resembling the natural appearance of an original scan. This template was constructed from a carefully chosen subset of 70 images from the Indian population-specific IBASD dataset, all acquired using the same scanner and parameters. The brain region was cropped, and standard zero-padding was applied to each image. Finally, the group registration tool from the ANTS package was used to align and combine the preprocessed images, creating the reference template. \\
\textit{Template Prepossessing} - The reference template image played a crucial role in normalizing each individual template across all four datasets. After applying cropping and standard zero-padding, affine alignment and histogram matching were conducted relative to the template using MATLAB, ensuring both affine and data acquisition invariance. To further ensure image quality and identify potential distortions, manual checks were performed during the preprocessing steps. Once verified, a time series of templates was generated for each population, consisting of six time points spanning ages from 20 to 80 years. This comprehensive approach allowed for robust and accurate analysis across different age groups in the study. \\
\textit{Creating Decade-Specific Templates} -
Finally, for standardization, the IBASD dataset was chosen as the reference population and a global template, combining all decade-specific templates, was created. Each individual template from the remaining three populations was affinely aligned and histogram equalized with respect to this global template. Visual checks  were performed in 3D to ensure the quality of each template before proceeding with further analysis. The preprocessing steps made the templates consistent and standardized across populations, shown in Figure \ref{BB} B. This ensures a reliable and accurate analysis of the aging process.
\subsection{An Overview of the Analytical Framework}
\begin{figure}[ht!]
    \centering
    \includegraphics[width=\linewidth]{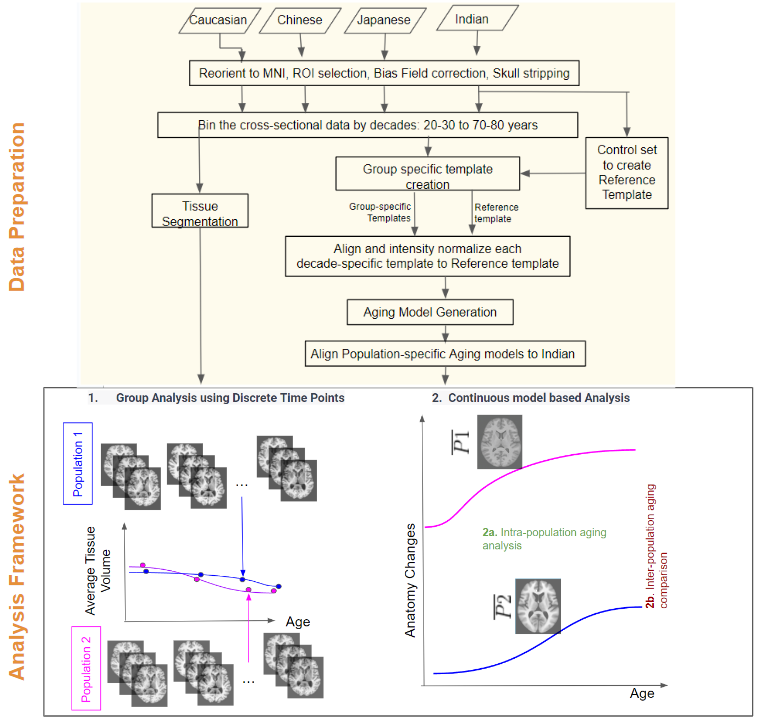}
    \caption{Graphical representation of analysis framework}
    \label{fr}
\end{figure}

Our aim is to investigate the structural variations in the aging brain in different populations. Both qualitative and quantitative measures are included in the analysis framework in order to understand population-specific aging processes. Figure \ref{fr} provides a visual representation of the proposed analysis framework. Global measures can be used to infer changes in brain size and tissue volume in different populations whereas non-rigid deformations can help understand the local changes. These local changes encompass both aging-related or time-dependent variations and population-specific, time-independent variations observed across populations.

\subsection{Group analysis using discrete time points}
In group analysis, the primary focus is indeed on examining the individual subject images to understand the group trend in aging for each population. 
Group analysis was chosen to be done via tissue volume changes across individuals. Tissue segmentation of the individual scans was first  performed using FSL-FAST. This is a tool in the FSL software package specifically designed for accurate tissue segmentation. Specifically, the grey matter, white matter, and CSF were segmented and the average volume of each segmented tissue was calculated for each decade by considering approximately 26 image tissue volumes at every time point. In order to visualise and capture the trend of volumetric changes, a cubic function was fitted to the mean volume points for each tissue type in each population. This trend allows us to quantitatively assess how tissue volumes change with aging and gain insights into the overall patterns and trajectories of tissue degeneration or growth.

\subsection{Continuous model based analysis}
Template-based analysis is a powerful method used to study aging-related changes by bringing individual images to a common space for cross-sectional data analysis. This approach involves aligning and registering the template images to a shared coordinate system. 
We choose a cross-sectional data-based diffeormorphic aging model (\cite{ours}) as it models the aging process as a smooth, monotonic, and diffeomorphic change observed across the discrete templates. The aging process is represented as a continuous deformation of a template in a global space. We next explain how this model is derived. \\
\textbf{Aging model}:\\
First, a template, represented as $S$, is obtained by aligning and then averaging all the templates in the discrete time series of a population. The aging process is then modeled as a deformation ($\phi$) of the average anatomy ($S$) over time. 
As $S$ represents a temporal average and is positioned towards the center in time, the aging deformation is defined with respect to this reference point. This design necessitates the computation of both forward and backward deformations relative to the reference time point. Given that $S$ is positioned towards the middle, $t=m$, two paths ($\phi_1$,$\phi_2$) are defined to cover the entire time range, namely $[t_0,m]$ and $[m,t_n]$. The forward and backward deformations are determined by composing the pairwise deformations sequentially. These deformations are computed between consecutive templates in the given time series and subsequently mapped into the $S$ space.

The aging model for the population is then found as
\begin{equation}
I(t)=\begin{cases}
 \phi_1(t) \circ S \mbox{ for } t\geq m, \\ 
  \phi_2(t) \circ S \mbox{ for } t\leq m. 
\end{cases}
\label{Tta}
\end{equation}
\subsubsection{Intra-population aging analysis}
The aging model given by Equation\ref{Tta} is used for this analysis. Visualizing the structural alterations in the brain aids in defining the expected trends of brain aging. Hence, an aging model is derived for each of the four populations (Indian, Chinese, Japanese and Caucasian). The normative deformations computed for each population are individually analyzed to gain insights into the local structural changes in the brain during aging. These deformations can be attributed to tissue contraction or expansion in fluid-filled regions as only topology preserving deformations are expected in a matured aging brain. Jacobians of the normative deformations help to draw meaningful conclusions.

\subsubsection{Inter-population aging analysis}
\label{dist}
The aging models utilized in this study are derived from different datasets, which introduces the possibility of temporal misalignment between the models. Additionally, the models may also exhibit affine misalignment and have varying intensity ranges. As mentioned earlier, these are addressed via normalisation and standardization processes which precede comparative analysis.
A two-level comparison framework is proposed aimed at providing a comprehensive understanding of the variations in aging among different populations. In the first level, a qualitative analysis is performed to explore trend variations by comparing the normative aging in individual populations. The differences in deformation trends for the entire brain and some specific sub-regions are visualised for this purpose.\\
Furthermore, deformation-based approaches are employed to explore the variations in brain structure and shape changes linked to the aging process across populations. To accomplish this, a metric described in our previous work \cite{ours2} is employed. In general, any difference in aging trends across different populations can be due to the difference between the average anatomies as well as the difference in deformation happening on the average anatomies. 
The former is termed global anatomy distance while the latter is termed as age dependant distance. Let us consider two populations $a$, $b$ with average/global anatomies $S_a$, and$S_b$, and  deformations $\phi^a$ and $\phi^b$, respectively. A deformation-based distance between the average anatomies ($S_a$ and $S_b$) quantifies the difference in the global anatomies while the maximum distance between the deformations ($\phi^a$ and $\phi^b$) in time quantifies the age dependant difference.
Specifically, the deformation between the $S_a$ and $S_b$ captures the global anatomy variation $\psi$ between the two time series; here, the deformation is modeled as  $\psi=\exp( \mathbf{V}_S)$, where $\mathbf{V}_S$ represents a stationary velocity field. Therefore, the norm of the vector field $\mathbf{V}_S$ can be directly used to quantify the deformation \cite{logdemon}. The global anatomy distance ($d_s$) between $S_a$ and $S_b$ is hence defined as
 \begin{equation}
     d_s=\left \| {\mathbf{V}_S} \right \|
     \label{shape}
 \end{equation}
The paths $\phi^a$ and $\phi^b$ are modeled with vector fields $v_a \cdot \gamma(t)$ and $v_b \cdot \gamma(t)$ respectively. The norm of the difference between the vector fields is used to compute the distance between the paths at each time point, $d_*(t)$. 
\begin{equation}
d_*(t)=\left \|v_a \cdot \gamma_a(t)-v_b \cdot \gamma_b(t) \right \|
\end{equation}
The age dependant distance is then defined to be the maximum $d_*(t)$.
\begin{equation}
    d_p=\max \left \{ d_{*}(t) \right \} 
    \label{age}
\end{equation}
It is important to note that both $d_S$ and $d_P$ are distances defined at every voxel of $S_a$ or $S_b$. By combining both qualitative and quantitative analyses, this framework offers a way to understand the variations in aging across populations.
\section{Results}
We illustrate the proposed framework using sample data collected for validation of the framework, highlighting the possibilities of conducting an elaborate analysis to derive various aspects of brain aging comparisons across populations.
\subsection{Group Analysis using Discrete Time Points}
\begin{figure}[ht!]
    \centering
    \includegraphics[width=\linewidth]{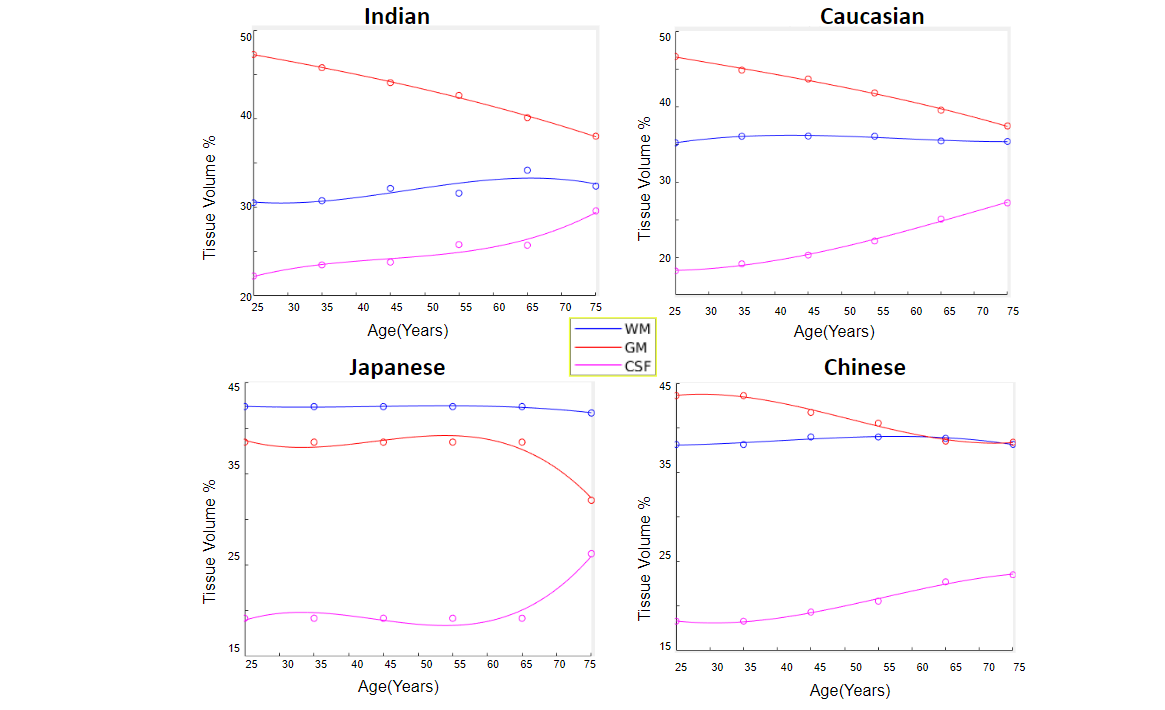}
    \caption{Distribution of White matter(WM), gray matter(GM), and cerebrospinal fluid (CSF) regions relative to the total brain volume across aging for different populations.}
    \label{intra_pop}
\end{figure}

An analysis was conducted to investigate the normative tissue volume changes relative to the total brain volume for different populations. The three major tissues, namely, white matter(WM), gray matter(GM), and cerebrospinal fluid (CSF), were analysed at individual scan level.
Segmented volumes were normalised with respect to the total brain volume in order to assess their proportion in the total brain volume. This was averaged over all scans at each time point. These are depicted for different populations in Figure \ref{intra_pop} with cubic spline-based interpolation between the time points. \\
In the following, all results refer to normalised values. When considering the gray matter (GM) volume,  it consistently decreases with age in Indian and Caucasian populations from earlier age points. In Japanese and Chinese groups the trend is different. Specifically, the Japanese population exhibits minimal GM degradation until the age of 60, after which there is a notable acceleration in degradation. In contrast, for the Chinese population, the onset of GM degradation occurs much earlier, starting at around 35 years of age, before which the degradation remains minimal. The WM volume tends to remain fairly stable across all populations except Indian which shows a slight deviation.
The plots also indicate ventricular expansion with aging for all populations as CSF volume has an increasing trend over age. However, it's worth noting that the acceleration of this expansion in the Japanese population occurs later, typically after the age of 65, which contrasts with other populations where the onset of ventricular expansion tends to occur earlier.
\subsection{Continuous model based Analysis}
\begin{figure}[ht!]
    \includegraphics[width=\linewidth]{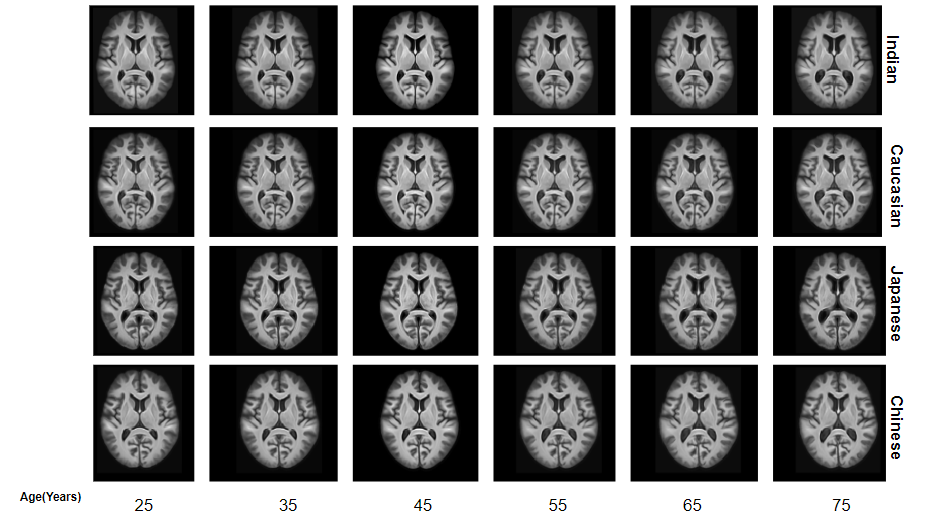}
    \caption{Age-specific templates derived from population-specific aging models for ages=25, 35, ...75 years}
    \label{intra_pop_mdl}
\end{figure}

Age-specific models were developed for each population using time series of brain templates. To create these aging models, templates were defined for each decade within the population. The Indian global template served as a reference to align other population templates using an affine transformation. The central slices of the age-specific templates derived from these aging models are shown in  Figure \ref{intra_pop_mdl}. As direct visual comparisons between the models is challenging, we performed further analyses to obtain interpretable results.

\subsubsection{Qualitative Analysis of Inter-Population Aging}
We wish to visualise the anatomical changes due to aging across various populations relative to a reference time point. On average, the brain is considered to be fully matured \cite{Johnson2009} at the age of 20 years and hence this serves as a good reference point.

\begin{figure}[ht!]
    \includegraphics[width=\linewidth]{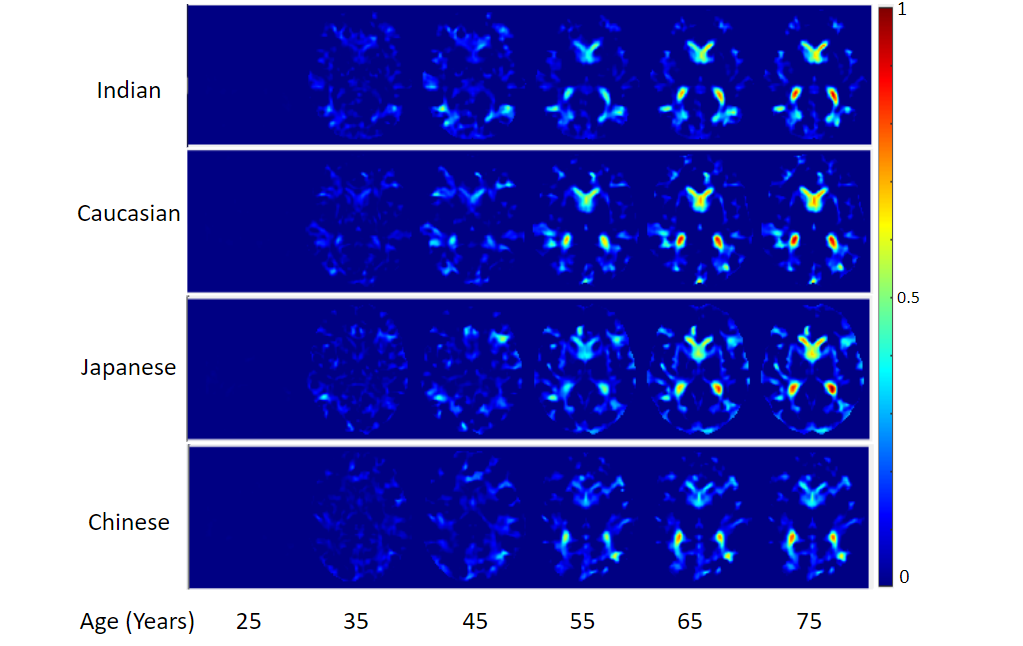}
    \caption{The deformation with respect to the initial time point (25years) and subsequent time points in each decade for every population}
    \label{intra_pop_d}
\end{figure}

\begin{figure}[ht!]
    \centering
    \includegraphics[width=\linewidth]{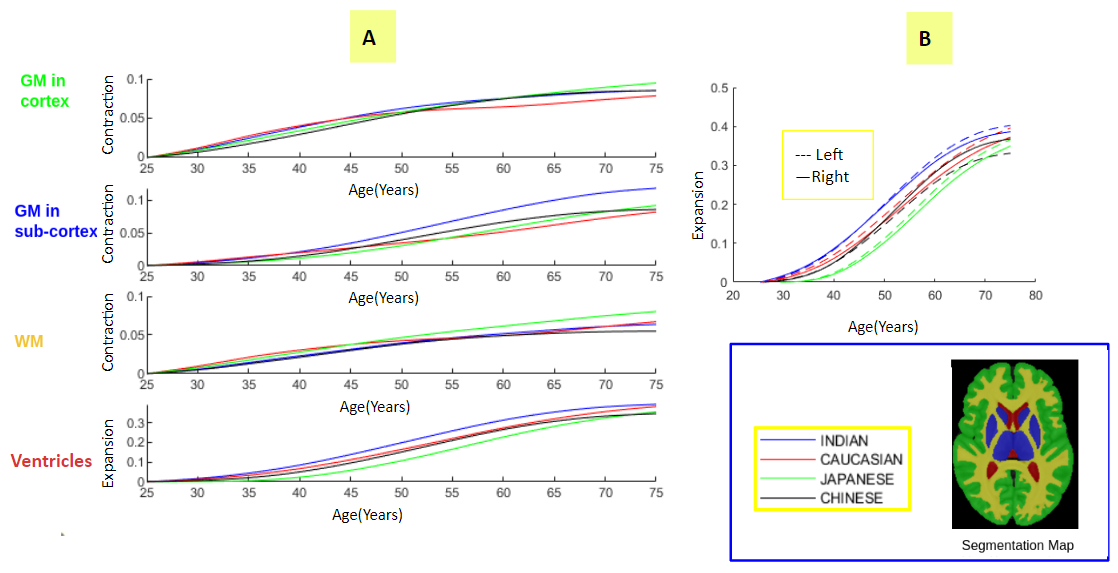}
    \caption{The plot of expansion and contraction in four brain regions over time, alongside the segmentation atlas. The regions include ventricles, sub-cortical structures, white matter, and cortical grey matter (shown inside the Blue box).}
    \label{intra_pop_p}
\end{figure}

The deformation between the brain anatomy and the reference brain from the same population was computed at every time point. Jacobians of the computed deformations were utilized to assess the local volume changes in different brain regions as they provide insights into the expansion or contraction occurring in specific brain regions. The resulting Jacobian maps, which illustrate the spatial distribution of these deformations, are visualised as heatmaps in Figure \ref{intra_pop_d}. The general trend observed is that there is no change at ages closer to 20 years (as indicated by blue pixels) and significant changes at ventricular locations above the age of 55. These maps enable an understanding of how the structural alterations in the brain evolve over time in different populations.
The tissue volume analysis plots (Figure \ref{intra_pop}) and the Jacobian maps (Figure \ref{intra_pop_d}) are correlated. 
In order to  see this more clearly, the rate of expansion or shrinkage in the brain was computed for specific regions using segmentation maps and their contraction/expansion relative to the reference point (20 years of age) were computed. Specifically, an atlas-based registration was used to identify four regions, encompassing ventricles, sub-cortical structures, white matter and cortical grey matter. Next, the Jacobians for these regions were computed and their average value at each time point was plotted as shown in Figure \ref{intra_pop_p} A. The ventricles predominantly exhibit increasingly positive Jacobian values over time, indicating an expanding trend, while the other regions show increasingly negative Jacobian values over time indicating shrinkage/contraction trend. There is evidence for early onset of contraction in cortical compared to sub-cortical GM. 
The sub-cortical gray matter contraction and ventricular expansion occurs at a significantly faster rate with an earlier onset in the Indian population when compared to other populations. Interestingly, white matter degradation is more pronounced in the Japanese population, with other populations exhibiting similar trends. The trends in cortical gray matter degradation become notably more distinct among populations after the age of approximately 55 years. Additionally, the analysis also highlights the delayed onset of ventricular expansion and sub-cortical contraction in the Japanese population.

An additional analysis was conducted to examine potential hemispherical asymmetry in brain aging. Figure \ref{intra_pop_p}-B) shows the average Jacobian values at the hemispherical level for the CSF-filled region in each population. The trends in the plots indicate that the expansion of this region in the left brain is at a much faster rate than in the right hemisphere and is consistently so in all populations.

\subsubsection{Quantitative analysis of inter-population aging}
\begin{figure}[ht!]
    \centering
    \includegraphics[width=\linewidth]{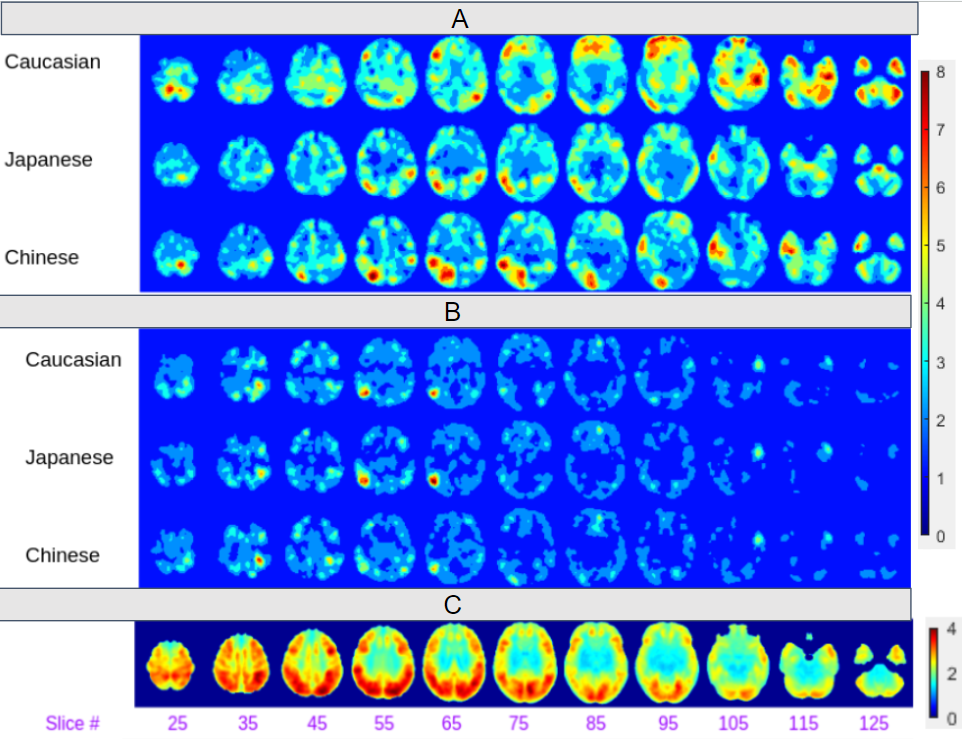}
    \caption{Brain aging differences across populations compared to the Indian population: (A) Global anatomy difference in 2D slices and (B) Age-dependent distance analysis. (C) Within-population deformations with respect to global anatomy for the Indian population. Left side correspond to Left side of Brain.}
    \label{inter_pop}
\end{figure}

It should be noted that all the previous analyses used brain tissue segmentations, which can introduce errors, particularly given the complexity of cortical tissue boundaries. In this analysis, structural changes are directly examined without utilizing any segmentation steps.
Here, we study the variations in average anatomies and deformations, via global anatomy and age-dependent distances, defined in Section \ref{dist}. Once again, the Indian brain is taken as a reference point. The global distance between the average anatomies were calculated using equation \ref{shape}, Since this distance is defined for every voxel we visualise the distance map as a heat map in Figure \ref{inter_pop} A. A red/yellow voxel indicates significant difference in the average anatomy for population X relative to that for Indian population.  Axial slices of the global distance map are shown in the figure. Global distance maps for the Chinese and Japanese are more similar, i.e., left Parietal-temporal-occipital regions has more variation. Whereas, the Caucasian distance map suggests more global anatomy mismatch towards the frontal region. \\
The pairwise age-dependent distance between Indian and another population was computed using Equation \ref{age}. These distances are visualised as heatmaps again in Figure \ref{inter_pop} B. From the predominance of blue pixels in the distance maps it is evident that, the aging trend between the Indian and other populations is similar except in some specific regions. The temporo-parietal regions however show higher age-dependant distances.\\
Since it is known that anatomical differences exist within a population, a baseline for population comparisons is useful. Hence, an analysis of intra-population deformations was conducted for the Indian population, with respect to its global anatomy. Each individual subject scan (spanning all age groups) was first registered to the global template image using a non-linear approach. Subsequently, the deformation-based distances i.e., the norm of the stationary velocity field parameterisation of the non-rigid deformation \cite{Yang2015} were calculated for each subject image from Global template. The average of all these distances is depicted in Figure \ref{inter_pop} C to illustrate the intra-population variation with respect to the Global template. It should be noted that the distance scale here is 0-4 whereas it is 0-8 for the maps in Figure \ref{inter_pop}A and B. Taking this scale difference into account, it is evident that inter-population differences are significantly more than the intra-population variations.
\begin{figure}[ht!]
    \centering
    \includegraphics[width=\linewidth]{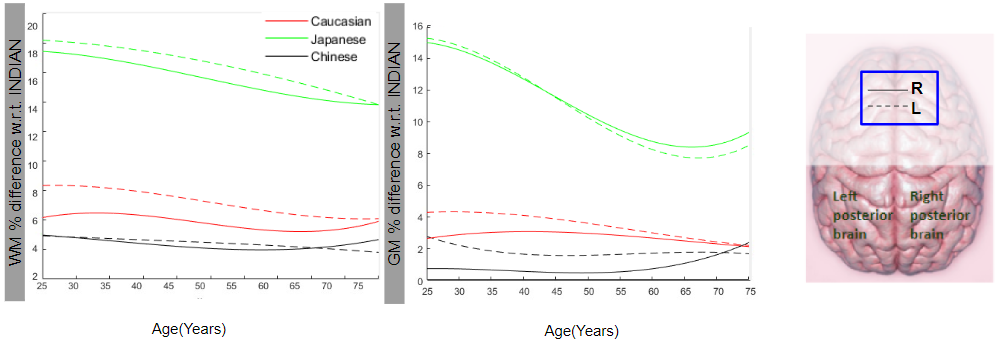}
    \caption{GM and WM contribution difference for different population with respect to Indian population for left and right posterior hemispheres}
    \label{kl}
\end{figure}
\\
Previously we observed that the posterior left and right hemispheres exhibited maximum age-dependant distances. To investigate this further, the absolute percentage differences in tissue volumes of a population with respect to Indian population were computed at every time point for the left and right posterior part of the brain. These are plotted in Figure \ref{kl} for GM and WM. In Figure \ref{kl}, the left posterior brain region specific plots have maximum slopes, which agrees with Figure \ref{inter_pop}C, where relatively maximum age-dependent changes are observed in the posterior region. Additionally, the left region showed comparatively more changes than the right.
Comparing the GM and WM plots in Figure \ref{kl} reveals that the Japanese population exhibits maximum tissue variation over time compared to the Indian population, primarily due to differences in GM.
\section{Discussion}
In this research, we adopted a systematic approach to analyze the neurological aging process across four distinct populations: Indian, Caucasian, Japanese, and Chinese, leveraging a dual-analytical strategy using in T1w structural MRI data. This is a preliminary attempt involving a detailed analysis across multiple populations. Rather than drawing definitive conclusions, we aim to discuss insights from the work for the given sample data, acknowledging the constraints of a smaller dataset size in this study. \\
The initial approach was based on a conventional group analysis utilizing tissue segmentation with FSL-FAST, showing volume changes across age in gray matter (GM), white matter (WM), and cerebrospinal fluid (CSF). Extending beyond this groundwork, we constructed a continuous model-based analysis to delve deeper into the intricate details of the aging process. Central to our approach was the representation of aging as a diffeomorphic transformation, a conceptualization that enabled a comprehensive intra- and inter-population analysis. This methodological choice allowed for a meticulous examination of both the overarching global anatomy and age-dependent variations, with a special focus on findings in reference to the Indian population. \\
A significant advancement of this study is the development of an analysis framework capable of comparing population-specific trends in aging both qualitatively and quantitatively by isolating the changes directly associated with aging. This innovation allows us to scrutinize voxel-level diffeomorphic transformations at a population level and discern differences across populations. Consequently, this study essentially underscores the importance of personalized approaches, opening avenues for potential clinical applications that are firmly rooted in a more population-specific and personalized understanding of neurological aging trajectories.\\
Our findings provide evidence of ventricular expansion and tissue degradation, consistent with previous research \cite{complongitudinal}. Notably, ventricular expansion accelerates within the age range of 40-45 years, and the results in Figure \ref{intra_pop_p} clearly indicate that the Indian population experiences an early onset of ventricular expansion compared to the other populations.Conversely, the Japanese population displays a delayed onset of ventricular expansion compared to other demographic groups. In the early stages, both the Chinese and Caucasian populations exhibit a similar pattern of ventricular expansion. However, in the elderly, the pattern diverges. The rate of expansion gradually stabilizes after approximately 70-75 years of age for the Chinese population, whereas the rate of expansion in the Caucasian population continues to increase, mirroring the trend observed in the other two populations. \\
Gray matter (GM) degeneration occurs at a faster rate than white matter (WM), which is consistent with the findings in \cite{brain_charts}. There were some interesting observations from our study results. The GM follows a linear pattern in both Caucasian and Indian populations in Figure \ref{intra_pop}. In the Chinese population, the degradation rate accelerates during the mid-age group and subsequently stabilizes in the elderly. Conversely, in the Japanese population, degradation commences at much later ages but then accelerates at a higher rate.  \\
When separately analyzing Cortical GM and sub-cortical GM contractions in Figure \ref{intra_pop_p}, it becomes evident that cortical GM contraction starts at earlier stages and follows similar trends across all populations, except for Caucasians, where the degradation rate is relatively lower in the elderly. Sub-cortical GM contraction, on the other hand, initiates around the mid-age range in all populations, with a relatively delayed onset observed in the Japanese population. Both cortical and sub-cortical GM tissue degradation increases with aging for all populations, but for the Chinese population, it subsequently stabilizes in the elderly. Japanese population has more white matter degradation compared to the other groups. A DTI-based investigation of WM changes across populations may be needed to confirm this observation. \\
Hemispherical asymmetry in aging is a well-established phenomenon \cite{assymminimal} and was observed in our study as well. In all populations, the left brain consistently exhibited faster ventricular expansion compared to the right brain, with a similar onset. The observed differences in slopes of the left and right ventricular expansion curves indicate that this asymmetry intensifies with aging in each population. Notably, the Chinese population displays more pronounced asymmetrical ventricular expansion with aging compared to other populations. In line with our findings, another study  \cite{CauvsChinese_assymetry} observes that Chinese individuals display more hemispherical asymmetry compared to Caucasians. Asymmetry in ventricular expansion is also observed in conditions such as Parkinson's disease \cite{asymmetrical}. Understanding population-specific ventricular expansion asymmetry can help define the normal limits tailored to each population and shed light on age-related changes and variations relevant to neurological conditions and the aging process. \\
Structural differences in the brain are prominent across populations, encompassing both global anatomical variances and age-related structural changes when compared to the Indian population. Global anatomical comparisons reveal that the left parietal-temporal-occipital regions exhibit more variation among Asian populations, while Caucasians show greater anatomical distinctions in the frontal region compared to the Indian population. The global anatomical differences are distributed throughout the brain, whereas age-related differences are relatively localized, with more pronounced deviations occurring in the parieto-temporal regions. These regions are recognized for their susceptibility to significant structural changes, particularly in neurodegenerative conditions such as Alzheimer's disease (AD). Performing automated segmentation-based statistical analysis of tissue changes for small regions can be error-prone, so multiple modalities, like susceptibility variation analysis functional changes etc. can help understand the desirable structural changes associated with aging in this region.
\\
This study is a pioneering effort in comprehending brain aging across diverse populations. It presents a detailed analysis framework that encompasses a comprehensive comparison of tissue degradation, hemispherical asymmetry development, age-related, and independent structural changes in brain anatomy across various populations. The analysis illustrated with sample data, and our intention
is to present this as a proof of concept to showcase the possibilities of analysis and potential insights using the proposed analysis framework. The same framework can be used for further exploration of critical factors like language, diet, and mental well-being, disease conditions underscoring the imperative for healthcare strategies tailored to the distinctive aging patterns found within each population. With the framework, the study is motivated to gain a deeper understanding of the underlying patterns and probable reasons for these aging trends. A larger dataset with more imaging modalities will be essential for future research.
\section{Conclusion}
In this study, a comprehensive comparative examination of brain aging is proposed, employing a novel analysis framework and demonstrated on a sample dataset encompassing four distinct populations, a pioneering effort in the field. Our examination encompassed Chinese, Caucasian, and Japanese populations, with the Indian population serving as our reference point. Through the comparative analysis, we observed variations in ventricular expansion, tissue degradation, and aging patterns across hemispheres, particularly noting structural differences in parieto-temporal brain regions.\\
The observations presented here are based on our framework applied to the sample dataset. While insightful, they shouldn't be overstated as standalone findings. Similar results can be expected with the proposed model, emphasizing its effectiveness. Further research is needed to validate and expand upon these initial findings, enhancing our understanding of age-related structural changes across populations. Importantly, the Japanese population showed significant aging deviations compared to Indians, with delayed onset aging and notable structural differences, particularly in tissue degradation and ventricular expansion. Both Japanese and Indian populations exhibited minimal hemispherical asymmetry compared to Chinese and Caucasians. Although Asian populations shared similar brain anatomies with Caucasians, differences in aging patterns highlight the usefulness of our framework for large dataset analysis.\\
These insights underscore the significance of redefining the concept of healthy aging tailored to each population, encouraging studies towards defining optimal normative aging for each population. Moreover, analyzing desired trends in aging across populations can aid in facilitating improvements in normative aging for others. These insights underscore the significance of redefining the concept of healthy aging tailored to each population, encouraging studies towards defining optimal normative aging for each population. Moreover, analyzing desired trends in aging across populations can aid in facilitating improvements in normative aging for others. Moreover, our framework aids in understanding the factors influencing aging through comprehensive comparisons, providing opportunities for interventions. Future research should explore factors like education and gender to better understand age-related structural changes across diverse populations.
\subsection* {Code, Data, and Materials Availability} 
 The datasets used and/or analysed during the current study available from the corresponding author on reasonable request.
\subsection* {Conflict of Interest} 
The authors declare no conflicts of interest.


\bibliography{report}   
\bibliographystyle{spiejour}   
\listoffigures
\listoftables

\end{spacing}
\end{document}